\documentclass{iopart}
\usepackage{iopams}
\usepackage{euscript,amssymb,epsfig}

\usepackage{amsfonts,latexsym}
\usepackage{euscript,amssymb}
\usepackage{epsfig}

\usepackage{graphicx}

\newcommand{\be}[1]{\begin{equation}\label{#1}}
\newcommand{\ee}{\end{equation}}
\newcommand{\ba}[1]{\begin{eqnarray}\label{#1}}
\newcommand{\ea}{\end{eqnarray}}
\newcommand{\rf}[1]{(\ref{#1})}
\newcommand{\nn}{\nonumber}

\begin{document}

\title{Kaluza-Klein models with spherical compactification: observational constraints and possible examples}

\author{Maxim Eingorn\dag\ddag \S}
\ead{maxim.eingorn@gmail.com}

\author{Seyed Hossein Fakhr\dag}
\ead{seyed.hossein.fakhr@gmail.com}

\author{Alexander Zhuk\dag}
\ead{ai.zhuk2@gmail.com}

\

\address{\dag Astronomical Observatory, Odessa National University,\\ Street Dvoryanskaya 2, Odessa 65082, Ukraine\\}

\

\address{\ddag Department of Theoretical and Experimental Nuclear Physics,\\ Odessa National Polytechnic University,\\ Shevchenko av. 1, Odessa 65044, Ukraine\\}

\

\address{\S Physics Department, North Carolina Central University,\\ Fayetteville st. 1801, Durham, North Carolina 27707, USA}

\

\begin{abstract}
We consider Kaluza-Klein models with background matter in the form of a multicomponent perfect fluid. This matter provides spherical compactification of the internal
space with an arbitrary number of dimensions. The gravitating source has the dust-like equation of state in the external/our space and an arbitrary equation of state
(with the parameter $\Omega$) in the internal space. In the single-component case, tension ($\Omega=-1/2$) is the necessary condition to satisfy both the gravitational
tests in the solar system and the thermodynamical observations. In the multicomponent case, we propose two models satisfying both of these observations. One of them also
requires tension $\Omega=-1/2$, but the second one is of special interest because is free of tension, i.e. $\Omega=0$. To get this result, we need to impose certain conditions.
\end{abstract}


\maketitle

\vspace{.5cm}


\section{\label{sec:1}Introduction}

\setcounter{equation}{0}

The modern theories of unification such as superstrings, supergravity and M-theory have the most self-consistent formulation in spacetime with extra dimensions.
Different aspects of the idea of multidimensionality are intensively used in numerous modern articles. Therefore, it is important to construct viable multidimensional
models which are in good agreement with physical experiments. The known gravitational experiments (the deflection of light and the time delay of radar echoes)
in the solar system are good filters to screen out non-physical theories. It is well known that the weak field
approximation is enough to calculate the corresponding formulas for these experiments \cite{Landau}. For example, in the case of general relativity these formulas
demonstrate excellent agreement with the experimental data.

In our previous papers  we investigated the popular Kaluza-Klein (KK) models with toroidal compactification of the internal spaces.  As we have shown, to be at the same
level of agreement with the gravitational tests as general relativity, the gravitating masses should have tension (i.e. negative pressure) in the internal spaces. This
is true for both linear \cite{EZ4,EZ5} and nonlinear $f(R)$ \cite{EZnonlin,EZf(R)solitons} models. At first glance, it looks unexpectedly, since dust-like equation of
state $p=0$ in all spatial dimensions is the most natural one  for nonrelativistic objects because the energy density for them is much bigger than the pressure. Such
approach for gravitating matter sources works very well in general relativity in the three-dimensional space to describe the gravitational tests in the solar system
\cite{Landau}. Therefore, we expected the same in Kaluza-Klein models. Unfortunately, the physically reasonable models with the dust-like equation of state $p=0$ in both
external and internal spaces contradict these tests \cite{EZ3}. It happens because of the fifth force generated by variations of the internal space volume \cite{EZ6}.
For the proper value of tension, the internal space volume is fixed and the fifth force is absent. It takes place, e.g., for the black strings/branes which have the
equation of state in the internal space $p=\Omega \varepsilon$ with $\Omega=-1/2$.

For black strings and black branes, the notion of tension is defined, e.g., in \cite{TF} and it follows from the first law for black hole spacetimes \cite{TZ,HO,TK}.
However, up to now we are not aware of the reasonable physical explanation for ordinary nonrelativistic objects, possessing such relativistic tension. Moreover, in the
recent paper \cite{ChEZ4} it was shown that in the case of non-zero tension there is a problem to formulate a many-body theory for such models. Black strings/branes are
compactified on tori. However, it is well known that superstring/supergravity models can be compactified also on Ricci-flat compact spaces (e.g., Calabi-Yau) and
spheres. For instance, 11-dimensional supergravity can be compactified on a torus $T^7$ and on a sphere $S^7$ (see, e.g., \cite{Wit,Nastase}). The Freund-Rubin mechanism
\cite{FR} is the most famous example of spontaneous compactification on $S^7$  where this is achieved with the help of the form fields. So, setting a goal to find viable
models without tension, we considered in \cite{ChEZ1,ChEZ2} Kaluza-Klein models where the internal space is a two-dimensional sphere and tension of the gravitating mass
is absent: $\Omega=0$. Here, we have shown that the conformal variation of the volume of the two-sphere generates the Yukawa-type admixture to the metric coefficients
and the non-relativistic gravitational potential. We estimated this admixture and found that it has a negligible value in the solar system. Therefore, this model (with
$\Omega=0$) satisfies the mentioned above gravitational tests. However, our subsequent research \cite{EZ7} indicates that tension ($\Omega=-1/2$) plays a crucial role
for the considered model because
the non-relativistic gravitating masses acquire effective relativistic pressure in the external space if $\Omega=0$. Obviously, such pressure contradicts the
observations. The equality $\Omega =-1/2$ is the only possibility to preserve the dust-like equation of state in the external space. So, we need tension again! How
general is this result? Is it possible, in principle, to construct a model without tension, satisfying the observations (from both gravitational and thermodynamical
points of view)? In the present paper we give an affirmative answer to this question and show what is the price for it.

Here, we consider a model where background matter is taken in the form of a multicomponent perfect fluid. This matter provides spherical compactification of the internal
space with arbitrary number of dimensions. The multicomponent approach is quite natural. For example, because of nontrivial topology of our multidimensional spacetime
(i.e. compactness of the internal space), vacuum fluctuations of quantized fields result in the nonzero energy density (the Casimir effect) \cite{Casimir}. Various
variants of the string theory contain real-valued form fields \cite{Stelle}. We mentioned above that these fields may result in spontaneous compactification due to the
Freund-Rubin mechanism. The model may also include other types of fluids. We perturb this background by a compact (usually, point-like) gravitating mass with the
dust-like equation of state in the external/our space and an arbitrary equation of state (with the constant parameter $\Omega$) in the internal space. In the
single-component case, we prove that to satisfy both the gravitational tests and thermodynamical observations, gravitating masses should have tension $\Omega=-1/2$.
However, in the multicomponent case, there is a possibility to construct a model which satisfies both the gravitational tests and the thermodynamical properties due to
multicomponent nature of the background matter. It takes place for any reasonable value of $\Omega$ including the most interesting dust-like case $\Omega=0$. However, to
achieve it we need a fine-tuning condition.

The paper is organized as follows. In Sec. 2 we define the background solution in the case of spherical compactification of the internal space with an arbitrary number
of dimensions. Here, the background matter takes the form of the multicomponent perfect fluid. Then, we perturb this solution by a gravitating mass with an arbitrary
equation of state in the internal space and obtain equations for the metric perturbations. In Sec. 3 we investigate in detail the single-component case and prove the
necessity of tension ($\Omega=-1/2$) for such model. Sec. 4 is devoted to the general multicomponent case. Here, we solve the equations for metric perturbations for two
particular examples and demonstrate a principal possibility for the considered model to satisfy the gravitational and thermodynamical observations in the case
$\Omega=0$. The main results are shortly summarized in the concluding Sec. 5.


\section{\label{sec:2}Multicomponent background solution and perturbations}

\setcounter{equation}{0}

Before the consideration of the gravitational field produced by the gravitating mass, we need to create an appropriate background metric. Such metric is defined on the
product manifold $M = M_4\times M_d$, where $M_4$ describes external four-dimensional flat spacetime and $M_d$ corresponds to the $d$-dimensional internal space which is
a sphere with the radius (the internal space scale factor) $a$, and should have the form
\be{2.1}
 ds^2=c^2 dt^2-dx^2-dy^2-dz^2 + \sum_{\mu=4}^D g_{\mu\mu}d\xi_{\mu}^2\, ,
\ee
where
\ba{2.2} g_{DD}=-a^2\, ,\quad g_{\mu\mu}=-a^2\prod_{\nu=\mu+1}^{D}\sin^2\xi_{\nu}\, ,\; \; \mu=4,\ldots ,D-1\, ,\ea
and $D=3+d$ is the total number of spatial dimensions. To create such metric with the curved internal space, we have to introduce background matter with the
energy-momentum tensor
\be{2.3}
\bar T_{ik}=\left\{
\begin{array}{cc}
\left( \frac{d(d-1)}{2\kappa a^2}-\Lambda_\mathcal{D} \right) g_{ik} & \mbox{for   } \, i,k=0,...,3;\\
\\
\left(\frac{(d-1)(d-2)}{2\kappa a^2} - \Lambda_\mathcal{D} \right)g_{ik} & \mbox{for   } \, i, k=4,5,\ldots ,D.
\end{array} \right . \quad
\ee
These components of the energy-momentum tensor can be easily got from the Einstein equation
\be{2.4} \kappa \bar T_{ik} = R_{ik}-\frac{1}{2} R g_{ik}-\kappa\Lambda_{\mathcal{D}} g_{ik}\,  \ee
for the background metric \rf{2.1}.
 Here, $\kappa \equiv 2S_D\tilde G_{\mathcal{D}}/c^4$, $S_D=2\pi^{D/2}/\Gamma (D/2)$ is the total solid angle (the surface area of the $(D-1)$-dimensional sphere of a
unit radius) and $\tilde G_{\mathcal{D}}$ is the gravitational constant in $(\mathcal{D}=D+1)$-dimensional spacetime. We also include in the model a bare
multidimensional cosmological constant $\Lambda_{\mathcal{D}}$. To get these components of the energy-momentum tensor, we took into account that the only nonzero
Ricci-tensor components are $R_{\mu\mu}=-[(d-1)/a^2]g_{\mu\mu},\; \mu=4,\ldots ,D$, and the scalar curvature is $R=-d(d-1)/a^2$. The minus sign in these formulas follows
from adopted here metric and curvature sign conventions (as in the book \cite{Landau}).

Now, we suppose that the energy-momentum tensor \rf{2.3} corresponds to the $N$-component perfect fluid:
\ba{2.5}
\bar T^0_{0}&=& \bar{\mathcal{E}}=\sum_{q=1}^{N}\bar\varepsilon_{q}\, ,\\
\label{2.6}\bar T^{\alpha}_{\alpha}&=&-\bar P_{0}=-\sum_{q=1}^{N}\bar
p_{(0)q}=-\sum_{q=1}^{N}\omega_{(0)q}\bar\varepsilon_{q}\, ,\quad \alpha = 1,2,3\, ,\\
\label{2.7}\bar T_{\mu}^{\mu}&=&-\bar P_{1}=-\sum_{q=1}^{N}\bar p_{(1)q}=-\sum_{q=1}^{N}\omega_{(1)q}\bar\varepsilon_{q}\, ,\quad\mu = 4,5,...,D\, , \ea
where $\omega_{(0)q}$  and  $\omega_{(1)q}$ are the parameters of equations of state of the $q$-th perfect fluid in the external and internal spaces, respectively.
It can be easily seen that the following relations take place:
\be{2.8}
\bar P_{0}=\left(-1\right)\bar{\mathcal{E}}\quad \Rightarrow\quad
\sum_{q=1}^{N}\left(1+\omega_{(0)q}\right)\bar\varepsilon_{q}=0\,
\ee
and
\ba{2.9}
\frac{\bar P_{1}}{\bar{\mathcal{E}}}&=&\frac{2\Lambda_{\mathcal{D}}\kappa a^{2}-\left(d-1\right)\left(d-2\right)}{d\left(d-1\right)-2\Lambda_{\mathcal{D}}\kappa a^{2}} \nn\\
&\Rightarrow& \sum_{q=1}^{N}\left[\omega_{(1)q}-\frac{2\Lambda_{\mathcal{D}}\kappa a^{2}-\left(d-1\right)\left(d-2\right)}{d\left(d-1\right)-2\Lambda_{\mathcal{D}}\kappa
a^{2}}\right]\bar\varepsilon_{q}=0 \, , \ea
or, equivalently,
\be{2.10}
\Lambda_{\mathcal{D}}=\frac{d-1}{\kappa a^2}\frac{\sum\limits_{q=1}^N\left(d
\omega_{(1)q}+d-2\right)\bar\varepsilon_q}{2\sum\limits_{q=1}^N\left(\omega_{(1)q}+1\right)\bar\varepsilon_q}\, .
\ee
There is also another useful relation:
\be{2.11}
\Lambda_{\mathcal{D}}=\frac12\sum\limits_{q=1}^N\left(d\omega_{(1)q}+d-2\right)\bar\varepsilon_q\, ,
\ee
which together with \rf{2.10} gives
\be{2.12}
\frac{d-1}{\kappa a^2}=\sum\limits_{q=1}^N\left(\omega_{(1)q}+1\right)\bar\varepsilon_q\, .
\ee
According to Eqs. \rf{2.10} and \rf{2.12}, in what follows we consider models that satisfy the inequality $\sum_{q=1}^N\left(\omega_{(1)q}+1\right)\bar\varepsilon_q \neq
0$.

Now, we perturb our background ansatz by a static point-like massive source with the nonrelativistic rest mass density $\hat \rho$. We suppose that the matter source is
uniformly smeared over the internal space \cite{EZ2}. Hence, multidimensional $\hat \rho$ and three-dimensional $\hat \rho_3$ rest mass densities are connected as
follows: $\hat \rho=\hat \rho_3({\bf r}_3)/V_{int}$ where $V_{int}=\left[2\pi^{(d+1)/2}/\Gamma((d+1)/2)\right]a^d$ is the surface area of  the $d$-dimensional sphere of
the radius $a$ (see, e.g., \cite{EZ3,EZ2}). In the case of a point-like mass $m$, $\hat \rho_3(r_3)=m\delta ({\bf r}_3)$, where $r_3=|{\bf r}_3|=\sqrt{x^2+y^2+z^2}$. In
the nonrelativistic approximation the energy density
of the point-like mass is $\hat{T}^0_{0}\approx \hat \rho c^2$ and up to linear in perturbations terms $\hat{T}_{00}\approx \hat \rho c^2$. Inasmuch as the gravitating
mass is at rest in the external space, it has the dust-like equation of state $\hat p_0=0\; \Rightarrow \; \hat{T}^1_{1}=\hat{T}^2_{2}=\hat{T}^3_{3}=0$ in our
dimensions. However, it may have the nonzero equation of state  $\hat p_1 \approx \Omega \hat \rho c^2\; \Rightarrow \; \hat T_{\mu}^{\mu}\approx -\Omega \hat \rho c^2
\, ,\; \mu =4,5,\ldots ,D$ in the internal space{\footnote{It is worth noting that, in KK models with toroidal compactification, the presence of nonzero pressure/tension
in the extra dimensions results in uniform smearing of the gravitating mass over the internal space \cite{ChEZ4}.}}. All other components of the energy-momentum tensor
of the gravitating mass are equal to zero.

Concerning the energy-momentum tensor of the background matter, we suppose that perturbation does not change the equations of state of the multicomponent perfect fluid
in the external and internal spaces, i.e. $\omega_{(0)q}$ and $\omega_{(1)q}$ are constants. Therefore, the energy-momentum tensor of the perturbed background is
\ba{2.13}
\tilde T_{00}&\approx& \left(\sum_{q=1}^{N}(\bar\varepsilon_{q}+\varepsilon_{q}^{1}) \right) g_{00}\, ,\\
\label{2.14}\tilde T_{\alpha\alpha}&\approx& -\left(\sum_{q=1}^{N}\omega_{(0)q}(\bar\varepsilon_{q}+\varepsilon_{q}^{1}) \right) g_{\alpha\alpha}\, ,\quad\alpha=1,2,3\, ,\\
\label{2.15} \tilde T_{\mu\mu}&\approx&-\left(\sum_{q=1}^{N}\omega_{(1)q}(\bar\varepsilon_{q}+\varepsilon_{q}^{1})\right) g_{\mu\mu}\, ,\quad\mu=4,5,...,D\, , \ea
where the corrections $\varepsilon^1_{q}$ are of the same order of magnitude as the perturbation $\hat \rho c^2$.

We suppose that the perturbed metric preserves its diagonal form. Obviously, the off-diagonal coefficients $g_{0\alpha}, \, \alpha =1,\ldots ,D$, are absent for the
static metric. It is also clear that in the case of the uniformly smeared (over the internal space) gravitating mass, the perturbed metric coefficients (see functions
$A,B,C,D$ and $G$ below) depend only on $x,y,z$ \cite{EZ2}, and the metric structure of the internal space does not change, i.e.
$g_{\mu\mu}=g_{DD}\prod_{\nu=\mu+1}^{D}\sin^2\xi_{\nu}\, , \mu=4,\ldots ,D-1$. The latter statement can be proved, e.g., in the weak field approximation from the
Einstein equations (see appendix B in \cite{ChEZ2}). It is also easy to show that in this case the spatial part of the external metric can be diagonalized by coordinate
transformations.
Therefore, the perturbed metric reads
\ba{2.16} \fl ds^2&=&Ac^2dt^2+Bdx^2+Cdy^2+Ddz^2+G\left[\sum_{\mu=4}^{D-1}\left( \prod_{\nu=\mu+1}^{D}\sin^2\xi_{\nu} \right)d\xi_{\mu}^2+d\xi_D^2\right] \ea
with
\ba{2.17}
&{}& A \approx 1+A^{1}({r}_3),\quad B \approx -1+B^{1}({r}_3),\quad C \approx -1+C^{1}({r}_3)\, ,\nn\\
&{}& D \approx -1+D^{1}({r}_3),\quad G \approx -a^2+G^{1}({r}_3)\, .\ea
All metric perturbations $A^1,B^1,C^1,D^1,G^1$ are of the order of quantities $\varepsilon^1_q$. To find these corrections as well as the background matter perturbations
$\varepsilon^1_q$, we should solve the Einstein equation
\be{2.18}
R_{ik}=\kappa\left( T_{ik}-\frac{1}{2+d}T g_{ik} -\frac{2}{2+d} \Lambda_{\mathcal{D}} g_{ik}\right)\, ,
\ee
where the energy-momentum tensor $T_{ik}$ is the sum of the perturbed background $\tilde T_{ik}$ \rf{2.13}-\rf{2.15} and the energy-momentum tensor of the perturbation
$\hat T_{ik}$. First, we would like to note that the diagonal components of the Ricci tensor for the metric \rf{2.16} up to linear
terms $A^1,B^1,C^1,D^1,G^1$ are
\ba{2.19}
R_{00} &\approx& \frac{1}{2}\triangle_3 A^1\, ,\nn \\
R_{11}&\approx&\frac{1}{2}\triangle_3 B^1
+\frac{1}{2}\left(-A^1-B^1+C^1+D^1+d\frac{G^1}{a^2}\right)_{xx}\, ,\nn \\
R_{22}&\approx&\frac{1}{2}\triangle_3 C^1
+\frac{1}{2}\left(-A^1+B^1-C^1+D^1+d\frac{G^1}{a^2}\right)_{yy}\, ,\nn \\
R_{33}&\approx&\frac{1}{2}\triangle_3 D^1
+\frac{1}{2}\left(-A^1+B^1+C^1-D^1+d\frac{G^1}{a^2}\right)_{zz}\, ,\nn \\
R_{DD}&\approx& d-1+\frac{1}{2}\triangle_3 G^1\, , \ea
where $\triangle_3=\partial^2/\partial x^2+\partial^2/\partial y^2+\partial^2/\partial z^2$ is the three-dimensional Laplace operator. Additionally, for the static
metric \rf{2.16}, where the coefficients $A,B,C,D$ and $G$ depend only on $x,y,z$, there is the following relation:
\be{2.20}
R_{\mu\mu} = R_{DD}\prod_{\nu=\mu+1}^{D}\sin^2\xi_{\nu}\, ,\quad \mu=4,\ldots ,D-1\, .
\ee
Concerning the off-diagonal components of the Ricci tensor, they should be equal to zero according to the Einstein equation \rf{2.18}. Taking into account the relations
$g_{\mu\mu}=g_{DD}\prod_{\nu=\mu+1}^{D}\sin^2\xi_{\nu}\, , \mu=4,\ldots ,D-1$, and that the metric coefficients $A,B,C,D$ and $G$ depend only on $x,y,z$, we can easily
verify that all off-diagonal components are identically equal to zero except the components $R_{12}, R_{13}$ and $R_{23}$. Equating these components to zero, we obtain
the following relations between the metric coefficients:
\be{2.21}
B^1=C^1=D^1
\ee
and
\be{2.22}
A^1-B^1-\frac{d}{a^2}G^1=0\, ,
\ee
which demonstrate that the expressions in brackets for 11, 22 and 33 components in \rf{2.19} vanish. Therefore, in the weak field limit the Einstein equation \rf{2.18}
is reduced to the following system of equations:
\ba{2.23}
&{}&\fl\triangle_3 A^1 =\frac{1+d+d\Omega}{1+d/2}\kappa\hat\rho c^{2} +\frac{\kappa}{1+d/2}\sum_{q=1}^{N}(d+1+3\omega_{(0)q}+d\omega_{(1)q})\varepsilon^{1}_q\, ,\\
&{}&\fl\triangle_3 B^1 =\frac{1-d\Omega}{1+d/2}\kappa\hat\rho c^{2}+\frac{\kappa}{1+d/2}\sum_{q=1}^{N}(1-\omega_{(0)q}+d\omega_{(0)q}-d\omega_{(1)q})\varepsilon_{q}^1\,
,
\label{2.24}\\
&{}&\fl\triangle_3 G^1 =\frac{1+2\Omega}{1+d/2}\kappa a^{2}\hat\rho c^{2}- \frac{2(d-1)}{a^2} G^1  +\frac{\kappa
a^2}{1+d/2}\sum\limits_{q=1}^N(1-3\omega_{(0)q}+2\omega_{(1)q})\varepsilon_q^1 \, ,\label{2.25}\ea
where we have used relations \rf{2.11} and \rf{2.12}. From these equations and the condition \rf{2.22} we obtain the connection between perturbations $G^1$ and
$\varepsilon^1_q$:
\be{2.26}
G^1=-\frac{2\kappa
a^4}{d(d-1)}\sum\limits_{q=1}^N\omega_{(0)q}\varepsilon_{q}^1\, .
\ee

We have four equations \rf{2.23}-\rf{2.26} to define $3+N$ unknown functions $A^1,B^1,G^1$ and $\varepsilon^1_q$. This is impossible in the general case $N>1$ without
some additional assumptions. However, for $N=1$, we can solve these equations exactly and this is the subject of the next section.

\section{\label{sec:3}Single-component background ($N=1$). Effective energy-momentum tensor for gravitating mass}

\setcounter{equation}{0}

Before we consider the general case $N>1$, it is instructive to explore in detail an exactly solvable single-component model $N=1$. For the single-component background,
it makes sense to drop the index $q$. Therefore, $\bar{\mathcal{E}}=\bar\varepsilon_1\equiv \bar\varepsilon$, $\bar P_0=\bar p_{(0)1}\equiv \bar p_0$, $\bar P_1=\bar
p_{(1)1}\equiv \bar p_1$, $\omega_{(0)1}\equiv \omega_0$ and $\omega_{(1)1}\equiv \omega_1$. Eq. \rf{2.8} demonstrates that the single-component background matter has
the vacuum-like equation of state in the external space:
\be{3.1a}
\bar{p}_0=\omega_0 \bar \varepsilon\, , \quad \omega_0 =-1,
\ee
but the equation of state in the internal space is not fixed:
\ba{3.2a}
&{}&\bar{p}_1=\omega_1\bar \varepsilon \, , \nn\\
&{}& \omega_1 = \frac{2\Lambda_{\mathcal{D}}\kappa a^2-(d-1)(d-2)}{d(d-1)-2\Lambda_{\mathcal{D}}\kappa a^2}\, \, ,
\ea
i.e. $\omega_1$ is arbitrary.  Choosing different values of $\omega_1$ (with fixed $\omega_0 = -1$), we can simulate different forms of matter. For example, $\omega_1=1$
corresponds to the monopole form-fields (the Freund-Rubin scheme of compactification \cite{FR})\footnote{In our case, they are $d$-forms (see, e.g., Eqs. (2.9) and (5.1)
in \cite{Zhuk}).}, and for the Casimir effect we have $\omega_1=4/d$ \cite{exci}. It is worth noting that the parameter $\omega_1$ can be positive only in the presence
of a positive bare cosmological constant $\Lambda_{\mathcal{D}}$. Moreover, it takes place only if $d-2<2\Lambda_{\mathcal{D}}\kappa a^2/(d-1)<d$. In contrast to the
model with the two-dimensional ($d=2$) sphere in \cite{ChEZ2}, the parameter $\omega_1$ does not disappear in the case of vanishing $\Lambda_{\mathcal{D}}$ for $d\geq
3$.

Taking into account that the relation \rf{2.26} has now the form
\be{3.3a}
\kappa\varepsilon^{1}=\frac{d(d-1)}{2  a^{4}}G^1\, ,
\ee
the system of Eqs. \rf{2.23}-\rf{2.25} reads:
\ba{3.4a}
&{}&\triangle_3 \left(A^1-\frac{d}{2a^2}G^1\right)=\kappa \hat\rho c^{2}=\frac{8\pi G_N}{c^2}\hat\rho_3\, ,\\
&{}&\triangle_3 \left(B^1+\frac{d}{2a^{2}}G^1\right)=\kappa \hat\rho c^{2}=\frac{8\pi G_N}{c^2}\hat\rho_3\label{3.5a}\, ,\\
&{}&\triangle_3 G^1 -\lambda^{-2}G^1=\frac{2(1+2\Omega)}{2+d}\kappa a^{2}\hat\rho c^{2} =a^2\frac{2(1+2\Omega)}{2+d}\frac{8\pi G_N}{c^2}\hat\rho_3\, ,\label{3.6a} \ea
where
\be{3.7a}
\lambda^2\equiv \frac{(d+2)a^2}{2(d-1)(d-2+d\omega_1)}
\ee
and we introduced the Newton gravitational constant
\be{3.8a}
4\pi G_N = \frac{S_D \tilde G_\mathcal{D}}{V_{int}}\, .
\ee
Let us consider now the point-like (in the external space) approximation for the gravitating objects: $\hat \rho_3(r_3)=m\delta ({\bf r}_3)$. The generalization of the
obtained results to the case of extended compact objects is obvious. It is well known that to get the physically reasonable solution of \rf{3.6a} with the boundary
condition $G^1 \to 0$ for $r_3 \to +\infty$, the parameter $\lambda^2$ should be positive, i.e. the equation of state parameter $\omega_1$ should satisfy the condition
\be{3.9a}
\omega_1> -1+\frac{2}{d}\, ,
\ee
which allows also negative values of $\omega_1$. From Eq. \rf{3.2a} we can get the corresponding restrictions for the bare cosmological constant:
\be{3.10a}
0<\frac{2\Lambda_{\mathcal{D}}\kappa a^2}{d-1}<d\, .
\ee
This inequality relaxes the condition of the positiveness of $\omega_1$. Then, the Eqs. \rf{3.4a}-\rf{3.6a} have solutions
\ba{3.11a}
A^1&=&\frac{2\varphi_N}{c^2}+\frac{d}{2a^{2}}G^1\, ,\\
B^1&=&\frac{2\varphi_N}{c^2}-\frac{d}{2a^{2}}G^1\, ,\label{3.12a}\\
G^1&=&a^2\frac{4\varphi_N}{(2+d)c^2}(1+2\Omega)\exp\left(-\frac{r_3}{\lambda}\right)\, ,\label{3.13a}
\ea
where the Newtonian potential $\varphi_N=-G_N m/r_3$. It is well known that the metric correction term $A^1\sim O\left(1/c^2\right)$ describes the non-relativistic
gravitational potential: $A^1=2\varphi/c^2$. Therefore, this potential acquires the Yukawa correction term:
\be{3.14a}
\varphi = \varphi_N\left[1+\frac{d}{2+d}(1+2\Omega)\exp\left(-\frac{r_3}{\lambda}\right)\right]\, . \ee

The inequalities \rf{3.9a} and \rf{3.10a} provide the condition of the internal space stabilization. Obviously, the monopole form-field ansatz with $\omega_1=1$
satisfies this condition. Let us consider this example in more detail. From the fine-tuning relation \rf{3.2a} we obtain $2\Lambda_{\mathcal{D}}\kappa a^2=(d-1)^2$.
Precisely this quantity provides in an effective dimensionally reduced model a zero value of the effective four-dimensional cosmological constant (see Eq. (5.7) in
\cite{Zhuk} where we should make the substitutions $\Lambda_{D} \to \Lambda_{\mathcal{D}}\kappa ,\, d_1 \to d ,\, D \to 4+d $ and where $\tilde R_1=d(d-1)/a^2,\, d_0=3,
D_0=4$). It is clear that in our case with flat background external spacetime, the effective four-dimensional cosmological constant should vanish. Additionally, this
value of $\Lambda_{\mathcal{D}}$ satisfies the stability condition (5.15) in \cite{Zhuk}. Moreover, the gravexciton/radion mass squared (5.12) (with substitution (5.11))
exactly coincides with the Yukawa mass squared $m_{Yu}^2=\lambda^{-2}=4(d-1)^2/[(d+2)a^2]\equiv m_{exci}^2$. Therefore, we arrived at the very natural and important
conclusion that the Yukawa mass is defined by the mass of the gravexcitons/radions.

It is of interest to estimate the Yukawa correction term in the formula \rf{3.14a}. It is worth mentioning that in our KK model, all gravitating masses (e.g., the balls
in the inverse square law experiment or the Sun and planets in the solar system) have in the non-relativistic limit the gravitational potential of the form \rf{3.14a}.
For reasonable values of the equation of state parameter $|\Omega|\sim O(1)$, the Yukawa parameter $\alpha$ (this is the standard notation for the prefactor in front of
the exponential function in the Yukawa potential) is also of the order of 1. Then, the inverse square law experiments restrict the characteristic range of the Yukawa
interaction \cite{new}: $\lambda \lesssim 10^{-3}$cm. Obviously, for the above-mentioned gravitational experiments in the solar system $r_3\gtrsim r_{\odot}\sim 7\times
10^{10}$cm, and the ratio $r_3/\lambda \gtrsim 10^{14}$. On the other hand, the collider experiments also restrict the sizes of the extra dimensions for Kaluza-Klein
models\footnote{We mean the KK models without branes. Precisely such models are considered in our paper.} \cite{AHN}: $a\lesssim 10^{-17}$cm. Since $\lambda \sim a$ (see
Eq. \rf{3.7a}), then for considered gravitational experiments in the solar system $r_3/\lambda \gtrsim 10^{28}$. It is clear that for such ratios we can drop the Yukawa
correction terms in Eqs. \rf{3.11a} and \rf{3.12a}. Therefore, the post-Newtonian parameter $\gamma=B^1/A^1$ is equal to 1 with extremely high accuracy for any value of
$\Omega$ including the most physically reasonable case of the dust-like value $\Omega=0$. The case $\Omega=-1/2$ is a special one, and we consider it below. Thus, our
model satisfies the gravitational tests (the deflection of light and the time delay of radar echoes) for any value of the equation of state parameter $\Omega$ in the
internal space.


So, at first glance, it seems that we have found a model, which, on the one hand, satisfies the gravitational experiments and, on the other hand, may not contain tension
(i.e. $\Omega =0$) in the internal space. However, let us examine in detail the energy-momentum tensor of the gravitating mass. As follows from Eqs. \rf{3.3a} and
\rf{3.13a}, the background perturbation $\varepsilon^1$ is localized around the gravitating object and falls exponentially with the distance $r_3$ from this mass.
Therefore, the bare gravitating mass is covered by this "coat".  For an external observer, this coated gravitating mass is characterized by the effective energy-momentum
tensor with the following nonzero components:
\ba{3.15a}
&{}&T_{0}^{0(eff)}\approx \varepsilon^1+\hat\rho({\bf r}_3)c^2=\nn \\
&{}&-(1+2\Omega)\frac{d(d-1)mc^2}{4(2+d)\pi V_{int}a^2}\frac{1}{r_3}\exp\left(-\frac{r_3}{\lambda}\right)+\frac{mc^2\delta({\bf
r}_3)}{V_{int}}\, ,\\
&{}&T_{\alpha}^{\alpha (eff)}\approx \varepsilon^1=\nn\\
&{}&-(1+2\Omega)\frac{d(d-1)mc^2}{4(2+d)\pi V_{int}a^2}\frac{1}{r_3}\exp\left(-\frac{r_3}{\lambda}\right),\quad\alpha=1,2,3\, ,\label{3.16a}\\
&{}&T_{\mu}^{\mu(eff)}\approx-\omega_1\varepsilon^1-\Omega\hat\rho({\bf r}_3) c^2=\nn\\
&{}&\omega_1(1+2\Omega)\frac{d(d-1)mc^2}{4(2+d)\pi
V_{int}a^2}\frac{1}{r_3}\exp\left(-\frac{r_3}{\lambda}\right)-\frac{\Omega mc^2\delta({\bf r}_3)}{V_{int}}\, ,\label{3.17a}\\
&{}&\mu=4,\ldots ,D\, ,\nn \ea
which, for illustrative purposes, can be presented as follows:
\ba{3.18a}
&{}&T_{0}^{0(eff)}\approx \frac{mc^2\delta({\bf r}_3)}{V_{int}}\left[1-\frac{d(1+2\Omega)}{2(d-2+d\omega_1)}\right]\, , \\
&{}&T_{\alpha}^{\alpha(eff)}\approx-\frac{mc^2\delta({\bf r}_3)}{V_{int}}\frac{d(1+2\Omega)}{2(d-2+d\omega_1)}\, ,\label{3.19a}\\
&{}&\alpha=1,2,3\, ,\nn\\
&{}&T_{\mu}^{\mu(eff)}\approx\frac{mc^2\delta({\bf r}_3)}{V_{int}}\, \frac{d\omega_1-2(d-2)\Omega}{2(d-2+d\omega_1)}\label{3.20a}\, ,\\
&{}&\mu=4,\ldots ,D\, ,\nn
\ea
where we have replaced the rapidly decreasing exponential function by the delta function:
\be{3.21a}
\frac{1}{r_3}\exp\left(-\frac{r_3}{\lambda}\right)\rightarrow\int\frac{1}{r'_3}\exp\left(-\frac{r'_3}{\lambda}\right)dV'_3 \times \delta({\bf r}_3)
=4\pi\lambda^2 \delta({\bf r}_3)\, . \ee
The less the parameter $\lambda$ is, the better this transformation is executed\footnote{\label{delta function}By definition of the delta-function
$$
\int f({\bf r}_3)\delta({\bf r}_3)d{\bf r}_3=f(0)\, ,
$$
that holds for any function $f({\bf r}_3)$ continuous at ${\bf r}_3=0$. It can be easily seen that
$$
\lim_{\lambda\to 0} \int f({\bf r}_3)\frac{1}{4\pi\lambda^2r_3}\exp\left(-\frac{r_3}{\lambda}\right)d{\bf r}_3=f(0)\, .
$$
Therefore,
$$
\lim_{\lambda\to 0} \frac{1}{4\pi\lambda^2r_3}\exp\left(-\frac{r_3}{\lambda}\right)=\delta({\bf r}_3)\, .
$$
}. On the other hand, the smaller the characteristic scale $\lambda$ of the Yukawa interaction is, the better concordance with the gravitational tests takes place for
our model. Obviously, taking into account the limitation $\lambda \lesssim 10^{-17}$cm and characteristic sizes of massive bodies, we can conclude that all corrections
connected with $G^1$ are negligible for the gravitational potentials outside of these bodies including atomic nuclei and up to astrophysical objects (see, e.g., Eqs.
\rf{3.11a}-\rf{3.13a}).  Unfortunately, this model faces the problem from the thermodynamical point of view, and this problem does not disappear, even in the case of
perfect concordance with the gravitational tests, i.e. for $\lambda \to 0$. The problem consists in the relativistic nature of $T^{\alpha (eff)}_{\alpha}$ in \rf{3.16a}
but not in the delta-function distribution, which we use for illustrative purposes. This equation (as well as Eq. \rf{3.19a}) demonstrates that the gravitating mass
acquires the effective relativistic pressure $\hat p_{0}^{(eff)}=-T_{\alpha}^{\alpha (eff)}$ in the external/our space. This relativistic component is concentrated in
the immediate vicinity of the point ${\bf r}_3=0$ and is carried along with the gravitating mass. This is the crucial point. To detect such pressure, we should place a
wall or a membrane in the path of a moving gravitating mass. The relativistic pressure should be felt at the moment when the mass crosses the wall or the membrane.
Clearly, it is hardly possible to carry out such experiment for the astrophysical objects. However, the formulas obtained above are suitable for any gravitating mass,
not only for astrophysical objects. For example, we can apply these expressions to a system of nonrelativistic particles forming, e.g., a gas of nonrelativistic
molecules. We can consider them as a system of point-like particles. This approach is well grounded in statistical physics (see, e.g., chapters I and X in
\cite{Klimontovich}) Then, as we have shown above, each of these particles is covered by the coat and this coat accompanies the moving particle. Let us put this system
in a box divided by a porous membrane. Obviously, Eqs. \rf{3.15a}-\rf{3.17a} (or Eqs. \rf{3.18a}-\rf{3.20a}) can be considered as ones in the co-moving frame with
respect to a particle. With the help of the Lorentz transformation, we can rewrite them in a laboratory system of coordinates (we shall denote it by the sign tilde)
connected with the box. If $v$ is a velocity of a molecule in the direction of the z axis ($\alpha =3$) which is perpendicular to the membrane, then 11 and 22 mixed
components remain the same (up to the trivial transformation of $r_3$) and the 33 component consists of two parts: $T_{\tilde 3}^{\tilde 3}=T_{\tilde 3}^{\tilde 3(eff)}
+ T_{\tilde 3}^{\tilde 3(bare)}$. The first part coincides with the 33 component in the co-moving frame: $T_{\tilde 3}^{\tilde 3(eff)}=T_{3}^{ 3(eff)}\sim mc^2$ up to
the transformation of $r_3$, i.e. it has the relativistic nature. The second part is connected with the non-relativistic motion of the particle: $T_{\tilde 3}^{\tilde
3(bare)} \sim m v^2$. These molecules move through the membrane to an empty part of the box.  It is well known that
momentum crossing the elementary spatial area $dx^{\tilde{\gamma}}\wedge dx^{\tilde{\delta}}$ per unit time is given by $T_{\tilde{\alpha}}^{\tilde{\beta}}
\varepsilon_{\tilde{\beta}|\tilde{\gamma}\tilde{\delta}|}dx^{\tilde{\gamma}}\wedge dx^{\tilde{\delta}}\, ,\; \tilde{\alpha},\tilde{\beta},\tilde{\gamma} = 1,2,3 $. In
the case of our membrane we get
\be{3.22a}
T^{\tilde 3}_{\tilde 3} d\tilde x\wedge d\tilde y = \left(T^{\tilde 3(eff)}_{\tilde 3}+T^{\tilde 3(bare)}_{\tilde 3}\right)d\tilde x\wedge d\tilde y\, .
\ee
Therefore, the non-relativistic particles have relativistic momentum crossing the membrane and resulting in relativistic pressure.

We can also get the relativistic pressure by averaging $T_{\alpha}^{\alpha (eff)}$ over a three-dimensional volume $V$:
\ba{3.23a}
\fl \overline{T}_{\alpha}^{\alpha (eff)}&\approx&-(1+2\Omega)\frac{d(d-1)mc^2}{4(2+d)\pi
V_{int}a^2}
\times\frac{1}{V}\int\frac{1}{r_3}\exp\left(-\frac{r_3}{\lambda}\right)dV\nn \\
\fl &\approx&
-(1+2\Omega)\frac{d(d-1)mc^2}{4(2+d)\pi
V_{int}a^2}\frac{4\pi\lambda^2}{V}
=-\frac{d(1+2\Omega)c^2}{2(d-2+d\omega_1)}\frac{m}{V_{int}V}\, ,
\ea
where we dropped the exponentially small terms. To get usual three-dimensional quantities, e.g., the pressure measured as $\mbox{erg/cm}^3$, we should multiply Eq.
\rf{3.23a} by $V_{int}$. Taking this remark into account and multiplying both sides of this equation by the total number $N$ of particles in the volume $V$, we obtain the
expression for the effective pressure of these particles in $V$:
\be{3.24a}
\hat P_0^{(eff)}\sim \rho c^2\, ,
\ee
where $\rho =Nm/V$ is the rest-mass density and we did not take into account the part of the pressure connected with non-relativistic motion of these particles. Of
course, such relativistic pressure contradicts the observations. It can be easily seen that the equality $\Omega =-1/2$ is the only possibility to achieve $\hat
p_{0}^{(eff)}=0$ for our model. It means that the bare gravitating mass should have tension with the equation of state $\hat p_1=-\hat \varepsilon /2$ in the internal
space. Then, the effective and bare energy densities coincide with each other and the gravitating mass remains pressureless in our space. In the internal space the
gravitating mass still has tension with the parameter of state $-1/2$. Therefore, to be in agreement with observations, the presence of tension is a necessary condition
for the considered model. However, we still do not know a physically reasonable explanation for the origin of relativistic tension ($\Omega=-1/2$) for non-relativistic
gravitating objects such as molecules, massive balls or our Sun. Moreover, as it was shown in the paper \cite{ChEZ4}, the presence of tension may result in difficulties
in the many-body problem for KK models.



\section{\label{sec:4}Multicomponent background $N>1$. Particular examples}

\setcounter{equation}{0}

Let us turn now to the general case $N>1$. As we have already mentioned at the end of Sec. 2, we need to assume some additional constraints to define perturbations
$A^1,B^1,G^1$ and $\varepsilon^1_q$ which satisfy Eqs. \rf{2.23}-\rf{2.26}. In this section, we provide two exactly solvable examples. Here, the additional constraint
(the fine-tuning condition) is chosen in such a way that, for the first example, the metric coefficients and the non-relativistic gravitational potential acquire
corrections in the form of the Yukawa potential, as in the previous section. Then, this example satisfies the gravitational tests in the solar system if the Yukawa
potential is negligible. For the second example, the fine-tuning condition will provide the condition of the constant/unperturbed internal space: $G^1=0$, as it takes
place for black string/branes with toroidal compactification. Such example satisfies the gravitational tests at the same level of accuracy as general relativity.

\subsection{Yukawa corrections}

In the first example, we generalize the solutions from the previous section, i.e. we shall find the Yukawa correction term to the nonrelativistic gravitational
potential. Above, we have shown that such term  arises in the case of stable compactification of the internal spaces (this is equivalent to the conditions \rf{3.9a} and
\rf{3.10a} in the single-component case), and the mass of the Yukawa interaction is defined by the mass of gravexciton/radion. A perfect fluid stabilizes the internal
space in the case of the vacuum-like equation of state in the external space \cite{EZ4,Zhuk}. Therefore, in this subsection we assume that
\be{3.1} \omega_{(0)q} = -1, \quad q=1,\ldots ,N\, . \ee
Note that with this choice of the parameters $\omega_{(0)q}$, Eq. \rf{2.8} is satisfied automatically.
Hence, Eq. \rf{2.26} is reduced to
\be{3.2}
G^1=\frac{2\kappa
a^4}{d(d-1)}\sum\limits_{q=1}^N\varepsilon_{q}^1\neq 0\, .
\ee
The case of the zero sum in \rf{3.2} (i.e. the zero perturbation $G^1$) will be considered in the next subsection. Additionally, we assume the following fine-tuning
condition:
\be{3.3}
\sum\limits^N_{\begin{array}{cc}
q,p=1\\
q\ne p
\end{array}}\omega_{(1)q}\varepsilon^1_p=0\, .
\ee
Together with the relation \rf{3.2}, this is equivalent to the following condition:
\be{3.4} \left(\sum\limits_{q=1}^N\omega_{(1)q}\right) G^1= \frac{2\kappa a^4}{d(d-1)} \sum\limits_{q=1}^N\omega_{(1)q}\varepsilon_{q}^1\, . \ee
Then, Eq. \rf{2.25} is exactly reduced to \rf{3.6a} where
\be{3.6} \lambda^{-2}\equiv \frac{1}{a^2(2+d)}\left[2(d-1)\left(d-2+d\sum\limits_{q=1}^N\omega_{(1)q}\right)\right] \ee
generalizes the definition \rf{3.7a}. The positiveness of $\lambda^2$ is the necessary condition of the internal space stabilization. For positive $\lambda^2$,
that takes place if the equation of state parameters $\omega_{(1)q}$ satisfy the
condition
\be{3.8}
\sum\limits_{q=1}^N\omega_{(1)q}> -1+\frac{2}{d}\, ,\quad d\geq 2\, ,
\ee
the solution $G^1$ has the form \rf{3.13a} (in the case of the point-like gravitating mass in the external space: $\hat \rho_3(r_3)=m\delta ({\bf r}_3)$).

It can be easily verified that, provided that the equations \rf{3.1}-\rf{3.4} hold, the equations \rf{2.23} and \rf{2.24} take the form of \rf{3.4a} and \rf{3.5a} with
solutions \rf{3.11a} and \rf{3.12a}. Therefore, the nonrelativistic gravitational potential $\varphi$ is given by the formula \rf{3.14a}, i.e. it has the Yukawa
correction term.

As a particular example, we consider the two-component model $(N=2)$ with the monopole form field $\omega_{(1)1}=1$ and vacuum fluctuations of quantized fields (Casimir
effect) $\omega_{(1)2}=4/d$ \cite{exci,Zhuk}. For these perfect fluids $\omega_{(0)1}=\omega_{(0)2}=-1$. Real-valued solitonic/monopole form fields naturally arise as
Ramond-Ramond form fields in the type II string theory and M-theory \cite{Stelle}. It is also well known that because of nontrivial topology of our multidimensional
spacetime (i.e. compactness of the internal space), vacuum fluctuations of quantized fields result in the nonzero energy density (the Casimir effect) \cite{Casimir}.
Therefore, from Eqs. \rf{3.2} and \rf{3.3} we get
\be{3.15} \varepsilon^1_1=-\frac{d}{4}\varepsilon^1_2=\frac{d^2(d-1)}{2(d-4)\kappa a^4} G^1\, . \ee
The case $d=4$ is excluded for this particular example because it contradicts the inequality \rf{3.2} $G^1\neq 0$.

We have shown in the previous section that the Yukawa correction terms to the metric coefficients are negligible for gravitating masses in the solar system.
Therefore, considered in this subsection examples satisfy the gravitational tests for any reasonable values $\Omega$ including the
most interesting dust-like case $\Omega=0$. Unfortunately, the nonrelativistic gravitating matter source acquires
effective relativistic pressure in the external/our space (remind that $\omega_{(0)q}=-1\, , \forall\,  q$):
\be{3.16} \hat p_0^{(eff)}=-T_{\alpha}^{\alpha (eff)}\approx -\sum\limits_{q=1}^N\varepsilon^1_q \sim m c^2\, , \quad \alpha = 1,2,3\, , \ee
which, certainly, is unacceptable from the thermodynamical point of view. The value $\Omega=-1/2$ is the only possibility to avoid it, i.e. the gravitating source should
have tension in the internal space. However, this is not the desired result. Therefore, in the next subsection we construct a model without tension, satisfying
observations from both gravitational and thermodynamical points of view.

\subsection{$\grave{\mbox{A}}$ la black brane with zero tension}

Let us assume now the following fine-tuning condition:
\be{3.17}
\sum_{q=1}^{N}\omega_{(0)q}\varepsilon_{q}^{1}=0\quad\Leftrightarrow\quad G^{1}=0\, ,
\ee
where the latter equation follows from Eq. \rf{2.26}. If all parameters of equations of state in the external space are equal to each other ($\omega_{(0)1}=\ldots
=\omega_{(0)N}$), then Eq. \rf{3.17} is reduced to
\be{3.18}
\sum_{q=1}^{N}\varepsilon_{q}^{1}=0\, ,
\ee
but we do not specially impose this condition. Taking into account Eq. \rf{3.17}, we get from \rf{2.25} the following additional relation:
\be{3.19}
\sum\limits_{q=1}^N\omega_{(1)q}\varepsilon_q^1=-\frac{1}{2}(1+2\Omega)\hat\rho c^2-\frac{1}{2}\sum\limits_{q=1}^N\varepsilon_q^1\, .
\ee
With the help of this relation and the condition \rf{3.17}, Eqs. \rf{2.23} and \rf{2.24} read
\ba{3.20}
\triangle_3 A^1&=& \kappa\hat \rho c^2+\kappa\sum_{q=1}^{N}\varepsilon_q^1\, ,\\
\label{3.21} \triangle_3 B^1 &=& \kappa\hat \rho c^2+\kappa\sum_{q=1}^{N}\varepsilon_q^1\, . \ea
We may conclude from these equations that metric coefficients $A^1=B^1$ for any value of $\Omega$ including the dust-like case $\Omega=0$. Therefore, the PPN parameter
$\gamma =1$ in full analogy with general relativity. So, we achieved in this model agreement with the gravitational tests (the deflection of light, the Shapiro effect)
in the solar system. To get the Newtonian limit, we should either consider the case \rf{3.18}, or assume that $\sum_{q=1}^N \varepsilon^1_q \sim \hat \rho c^2$. The
latter case results in renormalization of the multidimensional gravitational constant $\kappa$. The Newton gravitational constant is defined by Eq. \rf{3.8a} and the
nonrelativistic gravitational potential coincides exactly with the Newtonian expression:
\be{3.21}
A^1=\frac{2\varphi_N}{c^2}\, .
\ee

One of the main features of this model consists in the constant/unperturbed internal space because of $G^1=0$. It means that the conformal prefactor for the internal
space metric was not changed (up to $O(1/c^2)$): $G=-a^2=\mathrm{const}$. A similar situation takes place for black string/branes. Usually, they have the unperturbed
toroidal internal space. However, there is also generalization to unperturbed  spherical compactification \cite{ChEZ3}. For the considered in this subsection model, we
have shown (up to $O(1/c^2)$) that external spacetime metric is the Schwarzschild one and the internal space has the unperturbed spherical metric. The main advantage
of this model with respect to the black branes is a possibility to eliminate tension of the gravitating mass due to the dust-like choice $\Omega=0$. Because of this
difference we call this case "$\grave{\mbox{a}}$ la black brane".

This model is also satisfactory from the thermodynamical point of view because due to the condition \rf{3.17} the effective relativistic pressure (which is related to
the excitation of the background matter in Eq. \rf{2.14}) in the external/our space is absent:
\be{3.22}
\hat p_0^{(eff)}=-T_{\alpha}^{\alpha (eff)}\approx \sum\limits_{q=1}^N\omega_{(0)q}\varepsilon^1_q=0\, , \quad \alpha = 1,2,3\, .
\ee

As a particular example, we consider the two-component perfect fluid from the previous subsection where the monopole form field is characterized by the following
parameters of equations of state: $\omega_{(0)1}=-1,\, \omega_{(1)1}=1$, and for the Casimir effect we have $\omega_{(0)2}=-1,\, \omega_{(1)2}=4/d$. Hence, Eq. \rf{2.8}
is automatically satisfied. Then, for the most interesting dust-like case $\Omega =0$ we get from Eqs. \rf{3.18} and \rf{3.19} the following relations:
\be{3.23}
\varepsilon^1_1=-\varepsilon^1_2 = - \frac{d}{2(d-4)}\hat \rho c^2=- \frac{d}{2(d-4)}\frac{m c^2\delta({\bf r}_3)}{V_{int}}\, . \ee
The case $d=4$ is excluded for this particular example because it contradicts Eq. \rf{3.19} (here, $\omega_{(1)1}=\omega_{(1)2}$, and from Eq. \rf{3.19} we get a
non-physical result $\hat \rho =0$).

To conclude our investigations, we want to make the following remark. At the very end of Sec. 3, we wrote that we do not know a physically reasonable explanation for the
origin of relativistic tension ($\Omega=-1/2$) in the internal space for a non-relativistic gravitating object. However, our last example demonstrates how such an object
can acquire effective tension in the internal space. Let us consider the case where the bare parameter $\Omega$ for this gravitating mass is zero: $\Omega=0$.
Then, from Eqs. \rf{2.15} and \rf{3.19} we obtain the expression for the effective pressure of this object in the internal space:
\be{3.24}
\hat p_1^{(eff)}=-T_{\mu}^{\mu (eff)} \approx \sum_{q=1}^N \omega_{(1)q}\varepsilon^1_q =-\frac12 \hat\rho c^2\, ,\quad \mu =4,5,\ldots ,D\, ,
\ee
where we took into account the relation \rf{3.18}. That is we got an effective tension ($\Omega^{(eff)}=-1/2$). Hence, this is one more reason to call the considered
model $\grave{\mbox{a}}$ la black brane. It is worth noting that this effective tension arises not because of the bare tension of the gravitating mass, but because of
the perturbation of the background matter, and such procedure is physically well motivated.


\section{Conclusion}

In this paper, we investigated the viability of Kaluza-Klein models with spherical compactification of the internal space. To achieve such compactification, we
introduced background matter in the form of a multicomponent perfect fluid. The multicomponent approach is quite natural. For example, because of nontrivial topology of
our multidimensional spacetime (i.e. compactness of the internal space), vacuum fluctuations of quantized fields result in the nonzero energy density (the Casimir
effect). Various variants of the string theory contain real-valued form fields. The model may also include other types of fluids. This matter provides spherical
compactification of the internal space with an arbitrary number of dimensions $d$. We perturbed this background by a compact (in our three-dimensional space) gravitating
source with the dust-like equation of state in the external/our space $\hat p_0=0$ and an arbitrary equation of state $\hat p_1\approx \Omega\hat\rho c^2$ (with the
constant parameter $\Omega$) in the internal space. We assumed that this matter source is uniformly smeared over the internal space. In the weak-field limit (up to the
order $1/c^2$), we obtained equations for the perturbations of the metric and the background matter. It is impossible to solve these equations in the multicomponent case
$N>1$ because the number of equations is less than the number of unknowns. Hence, we need to introduce some additional fine-tuning relations.

However, a single-component case ($N=1$) is exactly solvable without additional constraints. Hence, we investigated this model in detail. The perturbed (up to
$O\left(1/c^2\right)$) metric coefficients were found from the Einstein equations. For the external space, these coefficients consist of two parts: the standard general
relativity expressions plus the admixture of the Yukawa interaction. The Yukawa interaction arises only in the case when the background matter satisfies some condition
(see the inequality \rf{3.9a}) for the parameter of the equation of state in the internal space which is equivalent to the condition of the internal space stabilization.
From the cosmological point of view, such stabilization was considered in papers \cite{exci,Zhuk} (see also the appendix in \cite{EZ4}). The stabilization takes place if
conformal excitations of the internal spaces (referred to as gravexcitons \cite{exci} or radions) acquire the positive mass squared. In our paper, we have got an
important result that the mass of the Yukawa interaction is exactly defined by the mass of the gravexciton/radion. In the solar system, the Yukawa mass is big enough for
dropping the admixture of this interaction and getting very good agreement with the gravitational tests for any value of $\Omega$.

Nevertheless, our subsequent investigation of this single-component model showed that the gravitating body acquires the effective relativistic pressure in the
external/our space. Clearly, it is hardly possible to detect such pressure for the astrophysical objects (we can not place a membrane or a wall in the path of these
objects), but any system of nonrelativistic particles such as molecules of a liquid or a gas may have the relativistic momentum crossing any spatial area. Of course, it
contradicts the observations. We have demonstrated that the value $\Omega=-1/2$ (i.e. tension!) is the only possibility to avoid this problem for the single-component
case.

It is worth noting that, in multidimensional models, the standard approach implies that to match the observations it is enough to stabilize the internal space. To
achieve it, the radion should have a relatively large mass. However, the single-component case demonstrates that this is not sufficient. In other words, the big enough
Yukawa/radion mass (for  models with the arbitrary physically reasonable parameter $\Omega$ in the equation of state in the internal space) is not sufficient to have a
good theory.  From the standard point of view, this result is fresh and unexpected.
The models should also possess the thermodynamical properties which do not contradict the observations. This natural demand leads to the unique value $\Omega=-1/2$ in
the case $N=1$, i.e. the gravitating mass should have tension in the internal space (the $d$-dimensional sphere). We have presented the detailed investigation of this
problem. Such investigation was absent before.

Then, we turned to the multicomponent case $N>1$.  Here, we need to assume some additional constraints (fine tuning conditions) to solve the Einstein equations. We
provided two exactly solvable examples.

In the first example (with the fine tuning condition \rf{3.3}), we arrived at the model with the Yukawa correction terms for the metric perturbations, by analogy with
the single-component case. For the sufficiently large Yukawa mass, this model satisfies the gravitational tests. Unfortunately, because of the same reasoning as in the
case $N=1$, the nonrelativistic gravitating matter source acquires the effective relativistic pressure in the external/our space which, certainly, is unacceptable from
the thermodynamical point of view. The value $\Omega=-1/2$ is the only possibility to avoid it, i.e. the gravitating source should have tension in the internal space.

In the next subsection we proposed the other model (with the fine tuning condition \rf{3.17}). Here,
the PPN parameter $\gamma$ is exactly equal to $1$ similar to general relativity. Therefore, this model satisfies the gravitational tests (the deflection of light, the
Shapiro effect) for any reasonable value of $\Omega$. Moreover, the nonrelativistic gravitational potential exactly coincides with the Newtonian one. One of the main
features of this model consists in the constant/unperturbed internal space. It means that the conformal prefactor for the internal space metric was not perturbed by the
gravitating mass. A similar situation takes place for black string/branes. The main advantage of this model with respect to the black branes is a possibility to
eliminate tension due to the dust-like choice $\Omega=0$. This model is also satisfactory from the thermodynamical point of view because the effective relativistic
pressure in the external/our space is absent: $\hat p_0^{(eff)}=0$. Therefore, we demonstrated a principal possibility for the considered model to satisfy the
gravitational and thermodynamical observations in the case of vanishing tension in the internal space for the gravitating mass. Certainly, it happens due to
multicomponent nature of the background matter. Although the fine-tuning conditions impose strong constraints on the model and look artificial, however we think that
they look less artificial than the bare relativistic tension ($\Omega=-1/2$). As an additional bonus for this model, we found a physically reasonable mechanism to
generate an effective tension for a gravitating mass. In this particular model, we demonstrated that a non-relativistic gravitating mass with zero bare tension
($\Omega=0$) acquires an effective tension in the internal space ($\Omega^{(eff)}=-1/2$)  due to the perturbation of the background matter, and such procedure is
physically well motivated.

\ack
This work was supported in part by the "Cosmomicrophysics-2" programme of the Physics and Astronomy Division of the National Academy of Sciences of Ukraine.
We want to
thank the referees for their valuable comments which have improved the presentation of the results.
A.~Zh.
acknowledges the hospitality of the Abdus Salam International Centre for Theoretical Physics and the Theory Division of CERN, where a part of this  work was carried out.
We also want to thank Prof. V.L. Kulinskii for useful discussions.


\end{document}